\documentclass[final]{elsart5p}
\usepackage{natbib,graphicx}
\usepackage{myfigures}
\journal{arXiv}

\begin{document}
\begin{frontmatter}
\title{Comprehensive structural classification of ligand binding motifs in proteins}  

\author[IPR]{Akira R. Kinjo\corauthref{cor}}
\corauth[cor]{Corresponding author}
\ead{akinjo@protein.osaka-u.ac.jp}
and
\author[IPR]{Haruki Nakamura}
\address[IPR]{Institute for Protein Research, Osaka University,
3-2 Yamadaoka, Suita, Osaka 565-0871, Japan}
\begin{abstract}
Comprehensive knowledge of protein-ligand interactions should provide
a useful basis for annotating protein functions, studying protein evolution,
engineering enzymatic activity, and designing drugs.
To investigate the diversity and universality of ligand binding sites in
protein structures, we conducted the all-against-all 
atomic-level structural comparison of over 180,000 ligand binding sites 
found in all the known structures in the Protein Data Bank by using a recently 
developed database search and alignment algorithm. 
By applying a hybrid top-down-bottom-up clustering analysis to the comparison 
results, we determined approximately 3000 well-defined structural motifs of 
ligand binding sites.
Apart from a handful of exceptions, most structural motifs
were found to be confined within single families or superfamilies, and 
to be associated with particular ligands.
Furthermore, we analyzed the components of the similarity network and 
enumerated more than 4000 pairs of ligand binding sites that were 
shared across different protein folds.
\end{abstract}
\end{frontmatter}
%%%%%
\section*{Introduction}
Most proteins function by interacting with other molecules.
Therefore, the knowledge of interactions between proteins and their 
ligands is central to our understanding of protein functions.
However, simply enumerating the interactions of individual proteins with 
individual ligands, which is now indeed possible owing to the massive 
production of experimentally determined protein structures, 
would only serve to increase the amount of data, not necessarily our knowledge
or understanding, of protein functions. 
What is needed is a classification of general patterns of interactions. 
Otherwise, it would be difficult to apply the wealth of information to 
elucidate the evolutionary history of protein functions \citep{AndreevaANDMurzin2006,Goldstein2008}, to engineer 
enzymatic activity \citep{GutteridgeANDThornton2005}, or to develop new drugs \citep{Rognan2007}. 

In order to classify protein-ligand interactions and to extract general 
patterns from the classification, it is a prerequisite to compare the 
ligand binding sites of different proteins. 
There are already a number of methods to compare the atomic structures or 
other structural features of functional sites of proteins 
 \citep[see reviews, ][]{JonesANDThornton2004,LeeETAL2007}.  

Applications of these methods lead to the discoveries of ligand binding site
structures shared by many proteins of different folds \citep{KobayashiANDGo1997,KinoshitaETAL1999,StarkETAL2003,BrakouliasANDJackson2004,Shulman-PelegETAL2004,GoldANDJackson2006}. 
\citet{GoldANDJackson2006} conducted an all-against-all comparison of 33,168 binding sites, the results of which have been compiled into the SitesBase database. They have described several 
unexpected similarities across different protein folds and applied their 
method to the annotation of unclassified proteins.
More recently, \citet{MinaiETAL2008} compared all pairs of 48,347 
potential ligand binding sites in 9708 representative protein chains, 
and demonstrated the applicability of ligand binding site comparison to 
drug discovery.

To date, however, no method has been applied to the exhaustive all-against-all 
comparison of all ligand binding sites found in the Protein Data Bank 
(PDB) \citep{wwPDB}, presumably because these methods were not efficient 
enough to handle the huge amount of data in the current PDB, 
or because it was assumed that the redundancy (in terms of sequence homology) 
or some ``trivial'' ligands (such as sulfate ions) in the PDB did not 
present any interesting findings. 
As of June, 2008, the PDB contains over 51,000 entries with more than 
180,000 ligand binding sites excluding water molecules, and hence naively 
comparing all the pairs of this many binding sites ($> 3\times 10^{10}$ pairs) 
is indeed a formidable task.
Nevertheless, multiple structures of many proteins that have been solved with 
a variety of ligands (e.g., inhibitors for enzymes) could provide a great opportunity 
for analyzing the diversity of binding modes, and some apparently trivial 
ligands are often used by crystallographers to infer the functional sites 
from the ``apo'' structure.
In other words, the diversity of these apparently redundant data is 
too precious a source of information to be ignored.

To handle this huge amount data, we have recently developed 
the GIRAF (Geometric Indexing with Refined Alignment Finder)
method \citep{KinjoANDNakamura2007}.
By combining ideas from geometric hashing \citep{GeometricHashing} and 
relational database searching \citep{DB_complete}, 
this method can efficiently find structurally and chemically similar 
local protein 
structures in a database and produce alignments at atomic resolution 
independent of sequence homology, sequence order, or protein fold.
In this method, we first compile a database of ligand binding sites into an 
ordinary relational database management system, and create an index based on 
the geometric features with surrounding atomic environments. 
Owing to the index, potentially similar ligand 
binding sites can be efficiently retrieved and unlikely hits are safely 
ignored. For each of the potential hits found, the refined atom-atom 
alignment is obtained by iterative applications of bipartite graph matching 
and optimal superposition. 
In this study, we have further improved the original GIRAF method 
so that one-against-all comparison takes effectively one second, and applied
it to the first all-against-all comparison of all ligand binding sites in 
the PDB.

In order to extract recurring patterns in ligand binding sites, 
we then classified the ligand binding sites based on the results of 
the all-against-all comparison, and defined structural motifs. 
So far, such structural motifs have been determined either 
manually \citep{PorterETAL2004} or automatically \citep{WangikarETAL2003,PolaccoANDBabbitt2006}. 
Given the huge amount of data, manual curation of all potential motifs is not 
feasible, and previously developed automatic methods are computationally 
too intensive \citep{WangikarETAL2003} or limited in scope (e.g., 
being based on sequence alignment \citep{PolaccoANDBabbitt2006}).
Therefore, we first applied divisive (top-down) hierarchical clustering 
to obtain single-linkage clusters from the similarity network of 
ligand binding sites which can be readily obtained from the result of 
the all-against-all comparison. 
Based on the hierarchy of the single-linkage clusters, 
agglomerative (bottom-up) complete-linkage clustering is then applied.
Thus obtained complete-linkage clusters are shown to be well-defined structural 
motifs, and are then subject to statistical characterization regarding 
their ligand specificity and protein folds. 

Furthermore, based on the result of the all-against-all comparison, 
we study the structure of the similarity network of ligand binding sites, 
and enumerate interesting similarities shared across different folds.
The list of clusters and the list of pairs of ligand binding sites not sharing 
the same fold are available on-line\footnote{http://pdbjs6.pdbj.org/$\sim$akinjo/lbs/}.
\section*{Results}
\label{sec:results}

\subsection*{All-against-all comparison of ligand binding sites}
\label{sec:allxall}

Out of 51,289 entries in the Protein Data Bank \citep{wwPDB} as of June 13, 2008,
all 186,485 ligand binding sites were extracted and compiled into a 
database. A ligand binding site is defined to be the set of protein atoms 
that are within 5\AA{} from any of the corresponding ligand atoms.
To define a ligand, we used the annotations in PDB's canonical XML 
(extensible markup language) files (PDBML) \citep{PDBML} because these 
annotations are more accurate than the HETATM record of 
the flat PDB files.
Our definition of ligands includes not only small molecules, but also 
polymers such as polydeoxyribonucleotide 
(DNA), polyribonucleotide (RNA), polysaccharides, and polypeptides 
with less than 25 amino acid residues; water molecules and ligands 
consisting of more than 1000 atoms were excluded. We did not exclude 
``trivial'' ligands such as sulfate ($\mathrm{SO}_{4}^{2-}$),
phosphate ($\mathrm{PO}_{4}^{3-}$), and metal ions.
We did not use a representative set of proteins based on 
sequence homology to reduce the data size. 

\begin{figure}[tb]
   \centering
\includegraphics[width=7cm]{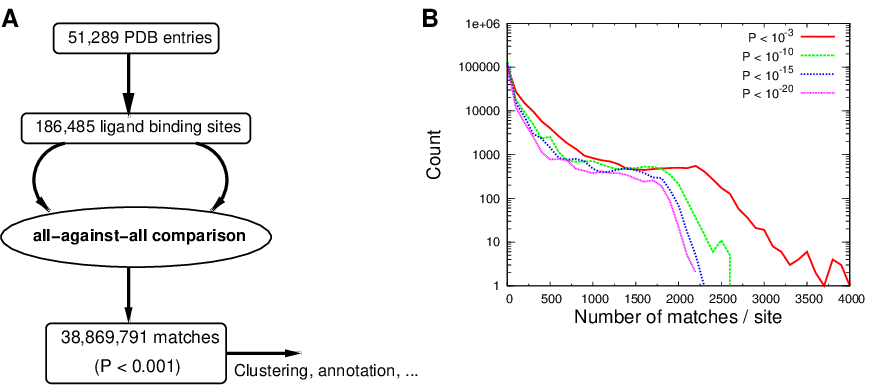}  
\caption{\figproclegend}
\end{figure}

In total, the all-against-all comparison yielded 38,869,791 matches with 
P-value $< 0.001$ with 208 matches per site on average (Fig. \figproc{}A). 
While 5014 sites found no hits other than themselves, 8369 sites found 
more than 1000 matches. When we limit the matches to more stringent P-value
thresholds ($10^{-10}$, $10^{-15}$, $10^{-20}$), the long tail of the large 
number of matches rapidly disappears (Fig. {\figproc}B), indicating 
that many matches reflect partial and weak similarities between sites.

\subsection*{Relationship between similarities of protein sequences and ligand binding sites}
\begin{figure*}[tb]
  \centering
\includegraphics[width=14cm]{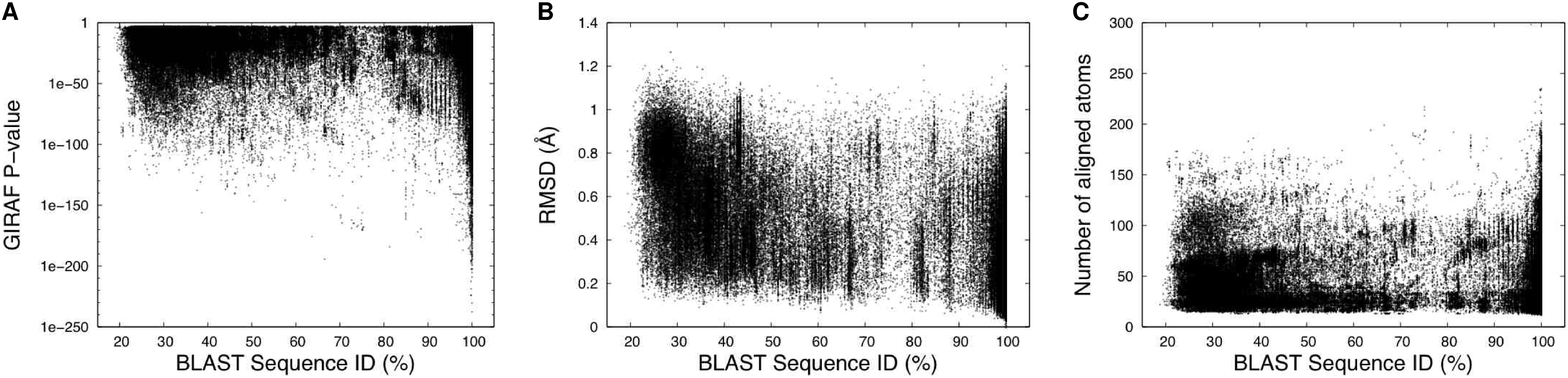}
  \caption{\figblastlegend}
\end{figure*}
As noted above, the present data set is highly redundant in terms of 
sequence homology. 
If the similarity of ligand binding sites is sharply correlated with that of 
amino acid sequences, it would have been better to use sequence representatives.
To justify the use of the redundant data set, we carried out an
all-against-all BLAST \citep{AltschulETAL1997} search of all protein chains 
of the present data set, and checked the correlation between sequence 
identity and GIRAF P-value (Fig. \figblast{}A).
It should be noted that a ligand binding site may reside at an interface 
of more than two protein subunits (chains), which complicates the notion of 
representative chains. Therefore, we defined sequence similarity between 
two PDB entries as the maximum sequence identity of all the possible pairs 
of chains from the two PDB entries. 

While there was a significant but 
very weak negative correlation between the GIRAF P-value and percent 
sequence identity (Pearson's correlation -0.14), there were many strikingly 
similar (GIRAF P-value $<10^{-50}$) pairs of ligand binding sites with low 
($<30$\%) sequence identity, and there were also many weakly similar ligand 
binding sites (GIRAF P-value $>10^{-20}$) at high ($>90$\%) sequence identity 
region. This tendency was also confirmed by using more conventional measures 
of similarities. Although the root-mean-square deviation (RMSD) of aligned 
atoms exhibited a stronger negative correlation with the sequence identity 
(Fig. \figblast{}B; Pearson's correlation -0.46), the range of scatter of RMSD 
was so large that it was not possible to distinguish the range of sequence 
identity from RMSD values and \emph{vice versa}. 
In addition, the number of aligned atoms did not 
correlate with the sequence identity (Fig. \figblast{}C), indicating that 
the local structures of ligand binding sites can be strictly conserved among
distantly related proteins. 
Visual inspection suggested a few possible reasons for the large 
deviation in the region of high sequence similarity. First, the binding sites 
do not necessarily overlap completely when different ligands are complexed 
with (almost) identical proteins. Second, many binding sites are flexible,
yet they are able to bind the same ligand. Third, some ligands are flexible and 
can be bound as different conformers, which in turn causes structural changes
of the binding site.

One of the rationales for an exhaustive all-against-all comparison is that some
similarities between non-representative proteins would be ignored when only
sequence representatives were used.
For example, in the results of a comparison of potential ligand binding 
sites of 9708 sequence representative proteins conducted by 
\citet{MinaiETAL2008}, the similarity between the ADP binding 
sites of human inositol (1,4,5)-triphosphate 3-kinase (PDB: 1W2D \citep{1W2D}; SAICAR synthase-like fold) and of \emph{Archaeoglobus fulgidus} Rio2 kinase (PDB: 1ZAR \citep{1ZAR}; Protein kinase-like fold)
was not detected although this match was found to have P-value of $8.1\times{}10^{-17}$ 
(40 aligned atoms; RMSD 0.75\AA{}) in the present result. Furthermore, 
equivalent matches were found in not all homologs of these two proteins. 
We note, however, that \citet{MinaiETAL2008} did find an equivalent similarity 
between 
the binding sites of these protein folds, but it was based on apo structures 
which were not treated here. Thus, the similarity not detected by Minai et al. 
is likely to be due to the use of representatives, but not due to the 
difference in sensitivity of their method and the present one.

We conclude that the similarity of sequences and that of ligand binding site 
structures are weakly correlated, but the correlation is not strong enough to
infer the one from the other.

\subsection*{Defining structural motifs of ligand binding sites}
\label{sec:cluster}
We have seen that sequence representatives are not suitable for studying
the diversity of ligand binding sites. The use of the raw data of ligand 
binding sites for statistical analysis, however, would be problematic due 
to some over-represented and under-represented binding sites. Therefore, it is 
preferable to remove the redundancy based on the ligand binding similarity 
itself. Furthermore, a list of pairwise similarities is not sufficient for 
characterizing typical patterns of binding modes.
Accordingly, we applied the hybrid top-down-bottom-up clustering method
to obtain complete-linkage clusters based on P-value.
In a complete-linkage cluster (hereafter referred to as `cluster'), 
any pair of its members are similar within the specified P-value 
threshold.
As such, clusters may be regarded as precisely defined structural motifs of 
ligand binding sites, and hence we use the term `cluster' and `structural 
motif' (or simply `motif') interchangeably when appropriate. 
Based on the analysis of similarity networks with varying thresholds 
(see below), we set the threshold to $10^{-15}$ in the following analysis.

\begin{figure*}[tb]
  \centering
\includegraphics*[width=14cm]{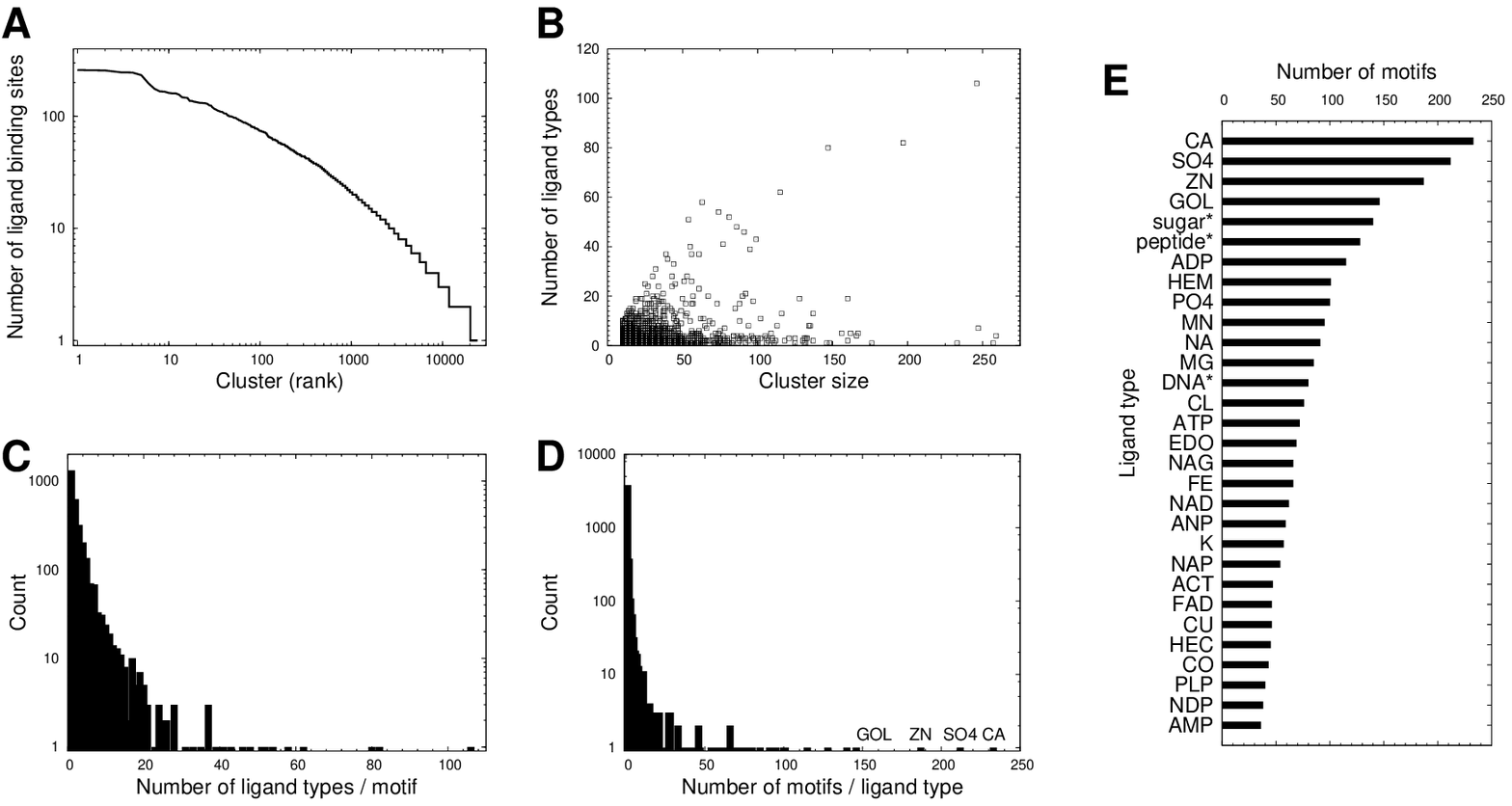}
  \caption{\figcclustlegend}
\end{figure*}
It is immediately evident that there are a large number of small clusters 
and a small number of large clusters (Fig. \figcclust{}A).
Excluding 58,001 singletons (clusters with only one member), 
there were 20,224 clusters which accounted for 
128,484 (69\%) of all the 186,485 sites.
Out of these clusters, 2959 clusters consisted of at least 10 sites, 
accounting for 69,748 (37\%) sites. The list of these clusters of 
structural motifs is available
on-line\footnote{http://pdbjs6.pdbj.org/$\sim$akinjo/lbs/cluster.xml}.
Since the ligand binding sites in small clusters are not reliable due to 
statistical errors, we use only the 2959 clusters consisting of at least 10  
sites in the following analysis unless otherwise stated.

\subsection*{Diversity of structural motifs with respect to ligand types}
\label{sec:divlig}

Although some structural motifs included binding sites for a wide variety 
of ligand 
types, this is not always the case (Fig. {\figcclust}B). 
Here, each PDB chemical component identifier (consisting of 1 to 3 letters) 
corresponds to a ligand type except for peptides, nucleic acids or sugars, 
which were treated simply as such (i.e., polymer sequence identity is ignored).
Large clusters associated with many kinds of ligands were almost always enzymes 
such as proteases (eukaryotic or retroviral), carbonic anhydrases, protein 
kinases and protein phosphatases, whose structures have been solved with 
a variety of inhibitors. 
For example, two structural motifs consisting of 246 and 147 ligand binding sites of 
eukaryotic (trypsin-like) proteases were associated with 106 and 80 ligand 
types, respectively; 
two motifs consisting of 197 and 115 sites of retroviral proteases 
with 82 and 62 ligand types, respectively;
a motif of 63 sites of protein kinases with 58 ligand types. 
On the contrary, large clusters with a limited variety of ligands were 
binding sites for heme (globins and nitric oxide synthase oxygenases) 
or metal ions. Each structural motif is associated with 3.2 ligand types on 
average (standard deviation 5.3): 
1322 motifs (47\%) with only one ligand type and 2807 motifs (95\%) 
with less than 10 ligand types whereas only 34 motifs contained 
more than 20 ligand types (Fig. {\figcclust}C). 
In general, the diversity of ligand types per structural motif is low.

The converse is also true. That is, the number of structural motifs 
associated with each ligand type is generally very limited with the average 
of 2.1 motifs (standard deviation 8.4) per ligand type (Fig. {\figcclust}D), 
and 3791 ligand types correspond to single motifs. 
Nevertheless, there were some ligands which 
were associated with many motifs (Fig. {\figcclust}E).
As expected, ligands often included in the solvent  
(e.g., SO4 [sulfate], MG [magnesium ion], GOL [glycerol], EDO [ethanediol])
were found in many motifs. Reflecting a large number of possible sequences, 
polymer molecules including peptide, sugar, and DNA were also found to be 
bound with many motifs, respectively. 
Other than these, mononucleotides and dinucleotides and metal ions exhibited
a wide range of binding modes. 

\subsection*{Diversity with respect to protein families and folds}
\label{sec:divfam}
Not many, but some structural motifs were found to contain ligand binding sites 
of distantly 
related proteins. To quantitatively analyze the diversity of structural motifs 
in terms of homologous families and global structural similarities, 
we assigned protein family, superfamily, fold and classes to each structural 
motif according to the SCOP \citep{SCOP} database. 
More concretely, the most specific SCOP code (SCOP concise 
classification string, SCCS) 
was assigned to each motif that was shared by all members of the corresponding 
cluster when it was possible, otherwise (i.e., there is at least one member
that is different from other members in the cluster at the class level),
motif was categorized as ``others'' (Fig. {\figcfold}A).

Out of 2705 motifs to which SCCS can be assigned, 2637 and 62 motifs 
shared the same domains at the family and superfamily level, respectively. 
Thus, more than 99\% of the motifs (of at least 10 binding sites) 
only contained binding sites of evolutionarily related proteins. 
One motif contained proteins from different superfamilies 
but of the same fold. This motif corresponded to the heme binding site of 
heme-binding four-helical bundle proteins (SCOP: f.21). 
Five motifs accommodated similarities across different folds,
out of which three were zinc binding motifs \citep{KrishnaETAL2003}. 
One motif contained a P-loop motif which is shared between the P-loop 
containing nucleotide triphosphate hydrolases (NTH) (SCOP: c.37) and 
the PEP carboxykinase-like fold (SCOP: c.91) (Fig. \figdifc{}A) \citep{TariETAL1996}.
One motif was of the nucleotide-binding sites from FAD/NAD(P)-binding domain 
(SCOP: c.3) and Nucleotide-binding domain (SCOP: c.4) (Fig. \figdifc{}B).
Note that some PDB entries have not yet been annotated in SCOP. Currently,
if such members exist in a cluster, they are simply ignored, and the assigned
SCCS is based only on the members whose SCCS is known. 
Therefore, the number of motifs not sharing the same folds is somewhat 
underestimated. Nevertheless, it seems a general tendency that most motifs 
are confined within homologous proteins, namely families or superfamilies.
\begin{figure}[tb]
  \centering
\includegraphics*[width=7cm]{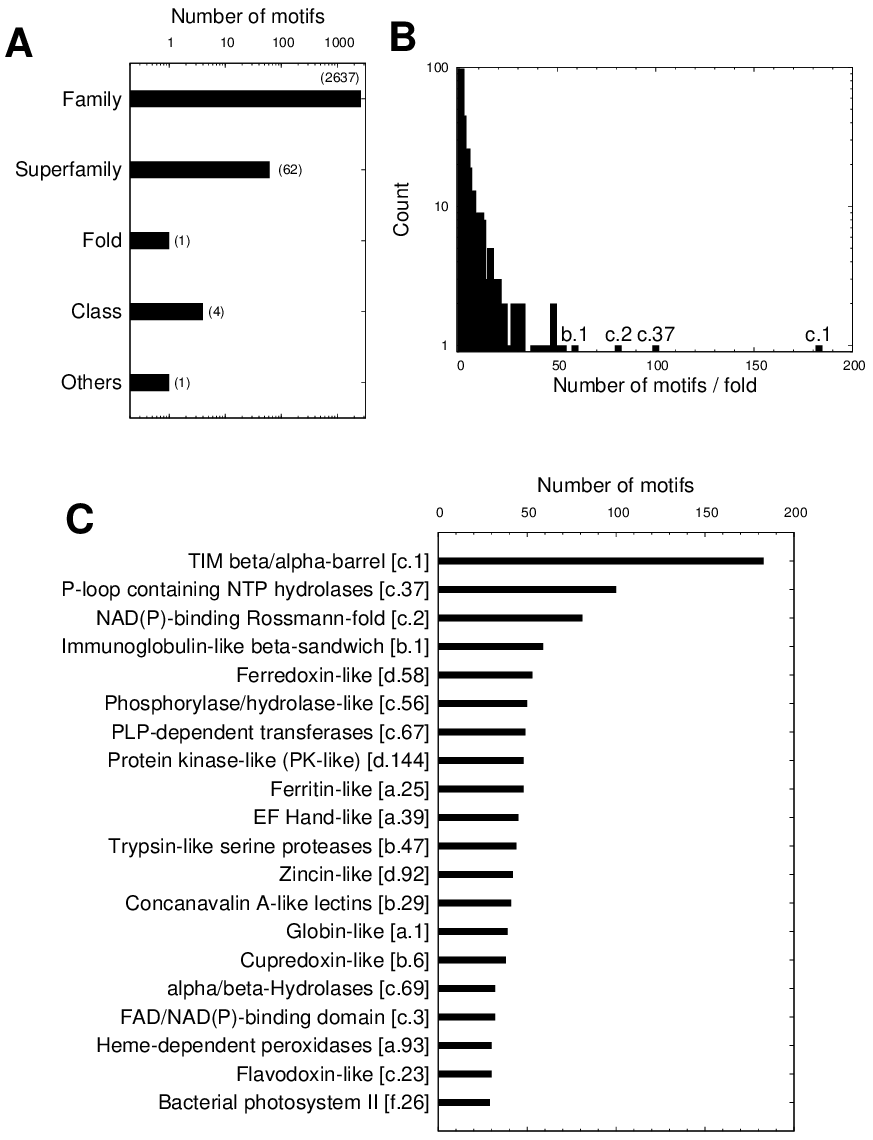}
  \caption{\figcfoldlegend}
\end{figure}

\begin{figure}[tb]
  \centering
\includegraphics[width=7cm]{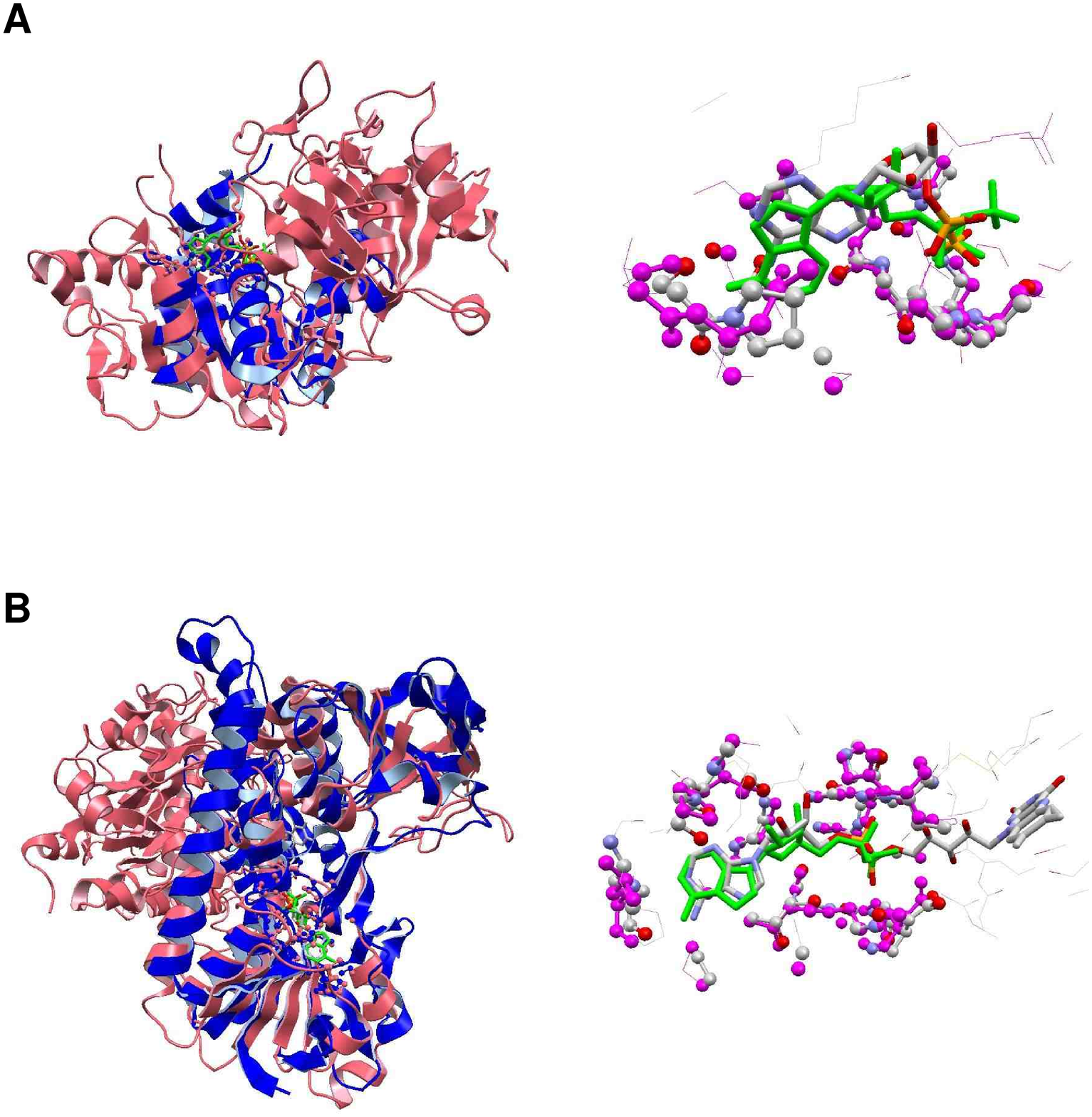}
  \caption{\figdifclegend}
\end{figure}
It was shown above that sequence similarity was only weakly related to
the structural similarity of ligand binding sites (Fig. \figblast{}).
This point can be further clarified by examining motifs of similar binding 
sites of related proteins.
For example, the peptide binding sites of a pig trypsin (PDB: 1UHB \citep{1UHB}) and 
of a human hepsin (PDB: 1Z8G \citep{1Z8G}) were both in the same cluster  
but they share little sequence similarity (5\% sequence identity based 
on a structural alignment \citep{KawabataANDNishikawa2000,Kawabata2003}), 
while the peptide binding site of bovine trypsin (PDB: 1QB1 \citep{1QB1}) 
in another cluster shares 81\% sequence identity with the pig thrombin 
in the previous cluster. 
This observation can be explained by the fact that different motifs cover 
different regions of proteins even though they are spatially close or even 
partially overlapping. The same argument applies to other motifs 
of related proteins.
Thus, the structural motifs distinguish subtle differences in ligand binding 
site structures independent of sequence similarity. 

It has been known that some protein folds can accommodate a wide range of 
functions. It is expected that the diversity of function is 
reflected in that of structures of ligand binding sites. 
To analyze such tendency, we counted the number of motifs that belong to 
each protein fold (Fig. {\figcfold}B). Only a handful of folds showed 
a large diversity in terms of structural motifs. On average, 8.9
motifs were assigned to a fold. Out of 332 folds used in the analysis, 
only 18 contained more than 30 motifs (Fig. {\figcfold}C). Among them, 
the TIM barrel fold was an extreme case with 183 motifs assigned, reflecting 
the great diversity of its functions \citep{NaganoETAL2002}.
Some superfolds \citep{OrengoETAL1994} such as Rossmann-fold, 
immunoglobulin-like, globin-like, etc. also showed great diversities of 
ligand binding sites.

\subsection*{Similarity network of ligand binding sites}
\label{sec:simnet}
While each motif defines a precise pattern of ligand binding mode, 
the members of different structural motifs share significant structural 
similarities with each other.
To explore the global structure of the `ligand binding site universe,' 
we constructed a similarity network based on the results of 
the all-against-all comparison. Each structural motif was represented 
as a node and two nodes were connected if a member of one node was  
significantly similar to a member of the other node (i.e., the P-value of 
their alignment was below a predefined threshold). 
Thus constructed network can be decomposed into 
a number of connected components. 
When the threshold was greater than $10^{-14}$, the size of the largest 
connected component of the network was one or two orders of magnitude 
greater than that of the second largest one (Fig. \fignetwork{}A).
For example, setting the threshold to $10^{-10}$ yielded the largest 
connected component consisting of 78,190 sites (i.e., 42\% of 186,485 sites).
Accordingly, many functionally unrelated binding sites were somehow 
connected in the largest component, which complicated the interpretation of
the component. 
With the P-value threshold of $10^{-15}$ or less, the first several 
connected components were of the same order (Fig. \fignetwork{}A), and 
many members of each component appeared to be more functionally related. 
Thus, we set P = $10^{-15}$ for constructing the network in the following
(as well as for defining the complete-linkage clusters described above).

Excluding 54,092 singleton components (those consisting of only one site), 
11,532 connected components were 
found. The largest component consisted of 7935 sites, and 1881 components 
contained at least 10 sites (Fig. \fignetwork{}A). 
\begin{figure*}[tb]
  \centering
\includegraphics*[width=12cm]{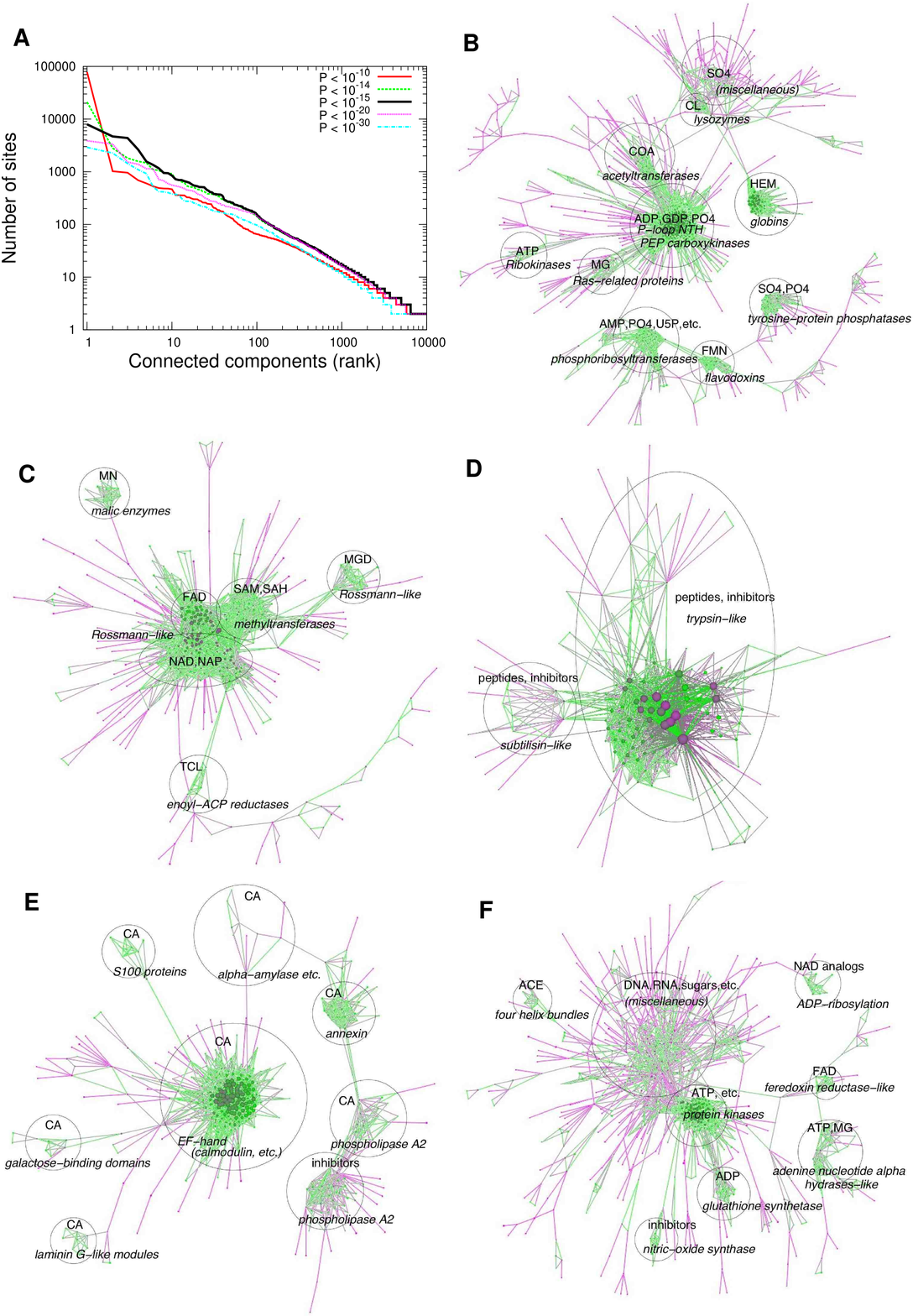}
  \caption{\fignetworklegend}
\end{figure*}

The main constituents of the largest connected component of the similarity 
network (Fig. \fignetwork{}B) were mononucleotide (ADP, GDP, etc.) or 
phosphate binding (PO4) sites.
Most notable were P-loop containing NTH (SCOP: c.37) and 
PEP carboxykinases (SCOP: c.91) which 
formed a closely connected group as they share similar phosphate binding 
sites, i.e., the P-loop motif (the term `group' used here indicates 
closely connected clusters in a network component colored in green in 
Fig. \fignetwork{}B-F). 
Directly connected with this group was the coenzyme A (CoA) binding site of 
acetyl-CoA acetyltransferases. 
The magnesium ion (MG) binding site of Ras-related proteins were also 
connected with the group of the P-loop containing proteins since the 
magnesium ion is often located near the phosphate binding site.
Mononucleotide or phosphate (AMP,U5P,PRP, PO4) binding sites of various 
phosphoribosyltransferases and the flavin mononucleotide (FMN) binding site 
of flavodoxins were also closely connected. The phosphate binding 
site of tyrosine-protein phosphatases formed another group which was 
weakly connected to the FMN binding site of flavodoxins. 

It is surprising that the heme binding site of globins 
(hemoglobins, myoglobins, cytoglobins, etc.) was also included in 
this component. Nevertheless, it was not directly connected 
to the main group of P-loops, but indirectly via the sparse group 
consisting of chloride ion binding site of T4 lysozymes and sulphate 
and phosphate binding sites of miscellaneous proteins. The binding sites 
of this latter group were made of regular structures at the termini 
of $\alpha$-helices. When we used a more stringent P-value threshold 
(say, $10^{-20}$), the groups of globins and lysozymes were detached from 
the main group, but the main group containing the P-loops was almost 
unaffected (data not shown). 
Thus, the matches connecting globins, lysozymes, and P-loop containing proteins
 may be considered as `false' hits. Based solely on structural similarity, 
however, they are difficult to discriminate from `true' hits 
(structural matches between functionally related sites) since many functional 
sites often include regular structures at termini of secondary structures. 
Nevertheless, the fact that only a subset of regular structures were detected 
suggests that these matches may correspond
to recurring structural patterns often used as building blocks of functional 
sites. In addition, we point out that weak but meaningful enzymatic functions 
are sometimes detected experimentally in such `false' hits \citep{IkuraETAL2008}.

The second largest connected component mainly consisted of 
mononucleotides or dinucleotides binding sites of the so-called 
Rossmann-like fold domains (Fig. \fignetwork{}C) which include, among others, 
NAD(P)-binding Rossmann-fold domains (SCOP: c.2), 
FAD/NAD(P)-binding domain (SCOP: c.3), nucleotide-binding domain (SCOP: c.4),
SAM-dependent methyltransferases (SCOP: c.66), activating enzymes of the ubiquitin-like proteins (SCOP: c.111) and urocanase (SCOP: e.51).

Peptide (and inhibitor) binding sites of trypsin-like and subtilisin-like 
proteases were found in the third largest component (Fig. \fignetwork{}D). 
These two proteases do not share a common fold, but were connected due to the 
similarity of the active site structures around the well-known catalytic triad.

The EF hand motif, a major calcium binding motif, was found in the fourth 
largest component (Fig. \fignetwork{}E) in which a variety of other calcium 
ion binding sites were also found. Although the main group in this component 
mainly consisted of the calcium ion binding sites of various calmodulin-like 
proteins, it also contained similar sites of periplasmic binding 
proteins (PBP). 
The ligands of these PBP's also include sodium in addition to calcium ions.
The main group was weakly connected to calcium ion binding sites of 
proteins of completely different folds such as galactose-binding domains 
(e.g., galactose oxidase, fucolectins), laminin G-like modules 
(e.g., laminin, agrin, etc.), alpha-amylases, annexins, and phospholipase A2. 
Due to its spatial proximity, the calcium binding site of phospholipase A2 
was also connected to its inhibitor binding sites.

Our last example, the fifth largest component, exhibited an exploding 
structure (Fig. \fignetwork{}F). Nevertheless, most binding sites are 
associated with nucleotides. The main closely connected group consisted of 
the ATP (and inhibitors) binding sites of protein kinase family proteins, 
next to which the ADP binding sites of glutathione synthetase family proteins
(including D-ala-D-ala ligases) were connected. Other closely connected groups
included FAD binding sites of ferredoxin reductase-like proteins, ATP or 
magnesium binding sites of adenine nucleotide alpha hydrolases-like proteins,
inhibitor binding sites of nitric-oxide synthases, and NAD (analog) binding 
sites of ADP-ribosylation proteins (e.g., T-cell ecto-ADP-ribosyltransferase 2,
iota toxin, etc.). There was a large sparse group connected 
with the main group of protein kinases. In that sparse group, ligand 
binding sites of transthyretins (prealbumins) were often found to be directly
connected with that of protein kinases although their folds are different.
These binding sites both involve a face of a $\beta$-sheet, and their 
similarity was found due to the backbone conformation of the $\beta$-sheet.
Since many proteins bind their ligands on a face of a $\beta$-sheet, 
this observation in turn explains the origin of the large sparse group.

\subsection*{Significant similarities across different folds}
\label{sec:diffold}
The similarity network of ligand binding sites revealed many structural 
similarities across different folds. To explore the extent of significant
`cross-fold' similarities (with $P< 10^{-15}$), 
we assigned SCOP codes to as many structural motifs as possible, 
and enumerated motif pairs whose members were significantly similar
 but did not share a common fold (Fig. \figdifstat{}A). 
We also examined the ligand pairs in those matches, and found that most of 
them were reasonable matches (Fig. \figdifstat{}B): metal ions were matched 
with metal ions, 
nucleotides with nucleotides or phosphate, and so on. Thus, many of 
these cross-fold similarities are expected to be functionally relevant.
The observation that sulfate (SO4) binding sites were often found to be 
matched with mononucleotide (GDP and ATP) or phosphate (PO4) binding sites 
(Fig. \figdifstat{}B) confirms the usefulness of the former ligand in inferring
the binding of the latter ligands, as often practiced by crystallographers.
We note that multiple SCCS may be assigned to a single motif if it 
contains multiple fold types or its member sites are located at an 
interface of multiple domains.
In order to cover all possible fold pairs, we did not exclude motifs 
consisting of less than 10 binding sites in this analysis.
There were in total 4035 pairs of structural motifs 
(52,709 pairs of binding sites) that exhibited significant similarities 
but did not share the same fold.
The complete list of these pairs is available 
on-line\footnote{http://pdbjs6.pdbj.org/$\sim$akinjo/lbs/diffold.xml}.
\begin{figure}[tb]
  \centering
\includegraphics*[width=7cm]{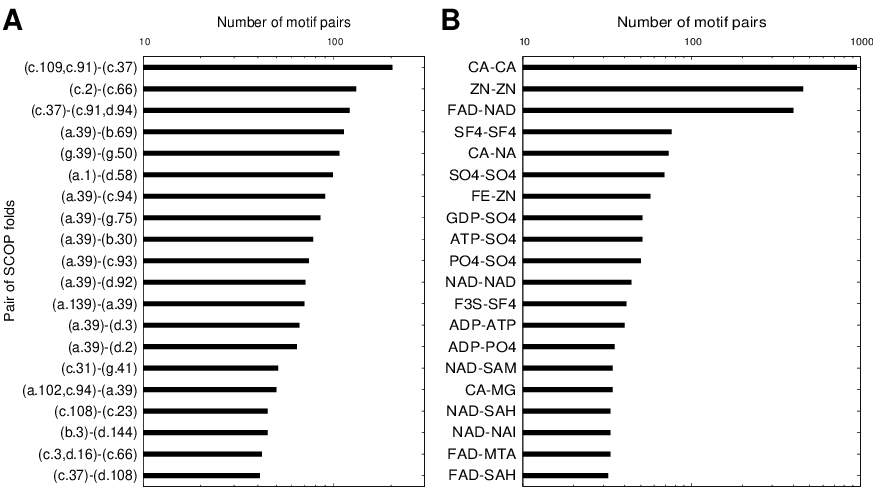}
  \caption{\figdifstatlegend}
\end{figure}

The most common cross-fold similarity was found between 
the P-loop containing NTH (SCOP: c.37) and the PEP-carboxykinase-like 
(SCOP: c.91) (c.f. Fig. \figdifc{}A). 
As described in the analysis of complete-linkage clusters, 
this corresponds to mononucleotide or phosphate binding sites. 

Mononucleotide or dinucleotide binding sites of various Rossmann-like folds 
(SCOP: c.2, c.3, c.4, c.66) also exhibited significant mutual similarities 
(e.g., Figs. \figdifc{}B and \figdiffold{}A). 

The calcium binding sites of EF hand-like fold (a.39) were found to be similar 
to the metal binding sites of many folds including beta-propeller proteins 
(Fig. \figdiffold{}B) and periplasmic binding proteins 
(SCOP: c.93 [class I], c.94 [class II]),
lysozyme-like (SCOP: d.2), Zincin-like (SCOP: d.92), and many others.

Similar zinc binding sites were found in many, mostly small, folds in addition 
to DHS-like NAD/FAD-binding domain (SCOP: c.31) and Rubredoxin-like (g.41)
(Fig. \figdiffold{}C), the former of which may be regarded as an inserted zinc 
finger motif.

The similarity between globin-like (SCOP: a.1) and 
ferredoxin-like (SCOP: d.58) was due to the coordinated structures of 
the iron-sulfur clusters found in alpha-helical ferredoxins and ferredoxins, 
respectively.

HAD-like fold proteins (SCOP: c.108) and CheY-like (flavodoxin fold) proteins 
(SCOP: c.23)  often share similar binding sites 
(e.g., Fig. \figdiffold{}D). Interestingly, although these proteins have 
very similar topologies, the orders of aligned secondary 
structure elements were different when the alignment was based on the ligand 
binding site similarity.

As noted in the description of a network component (Fig. \fignetwork{}F), 
protein kinases and transthyretins share similar binding sites which are 
located on a face of a $\beta$-sheet (Fig. \figdiffold{}E). Nevertheless,
their ligand moieties seem also similar.

Also as seen in the network component (Fig. \fignetwork{}B), phosphate 
binding site of the P-loop motif exhibits a significant similarity with CoA
 binding site of acetyltransferases (Fig. \figdiffold{}F). A close examination 
showed the phosphate bound to the P-loop motif coincided with the phosphate 
group of CoA bound to the acetyltransferase.

The list of the cross-fold similarities contained many other examples including,
but not limited to, those discussed in the context of the similarity network.
Here we give two other examples. Bacterial peptide deformylase 2 (SCOP: d.167) 
and human macrophage metalloelastase (SCOP: d.92) both act with peptides, and 
their ligand binding sites exhibit high structural similarity 
(Fig. \figdiffold{}G). DNA is one of the most abundant ligands found in cross-fold similarities 
(Fig. \figdifstat{}B). Not surprisingly, there can be also found similarity 
between binding sites for DNA and RNA.  One example is the KH1 domain of
human poly(rC)-binding protein 2 which binds DNA and bacterial transcription elongation protein NusA which binds RNA (Fig. \figdiffold{}H). These proteins have 
different variants of the KH domains \citep{Grishin2001}.
\begin{figure*}[tb]
  \centering
\includegraphics*[width=12cm]{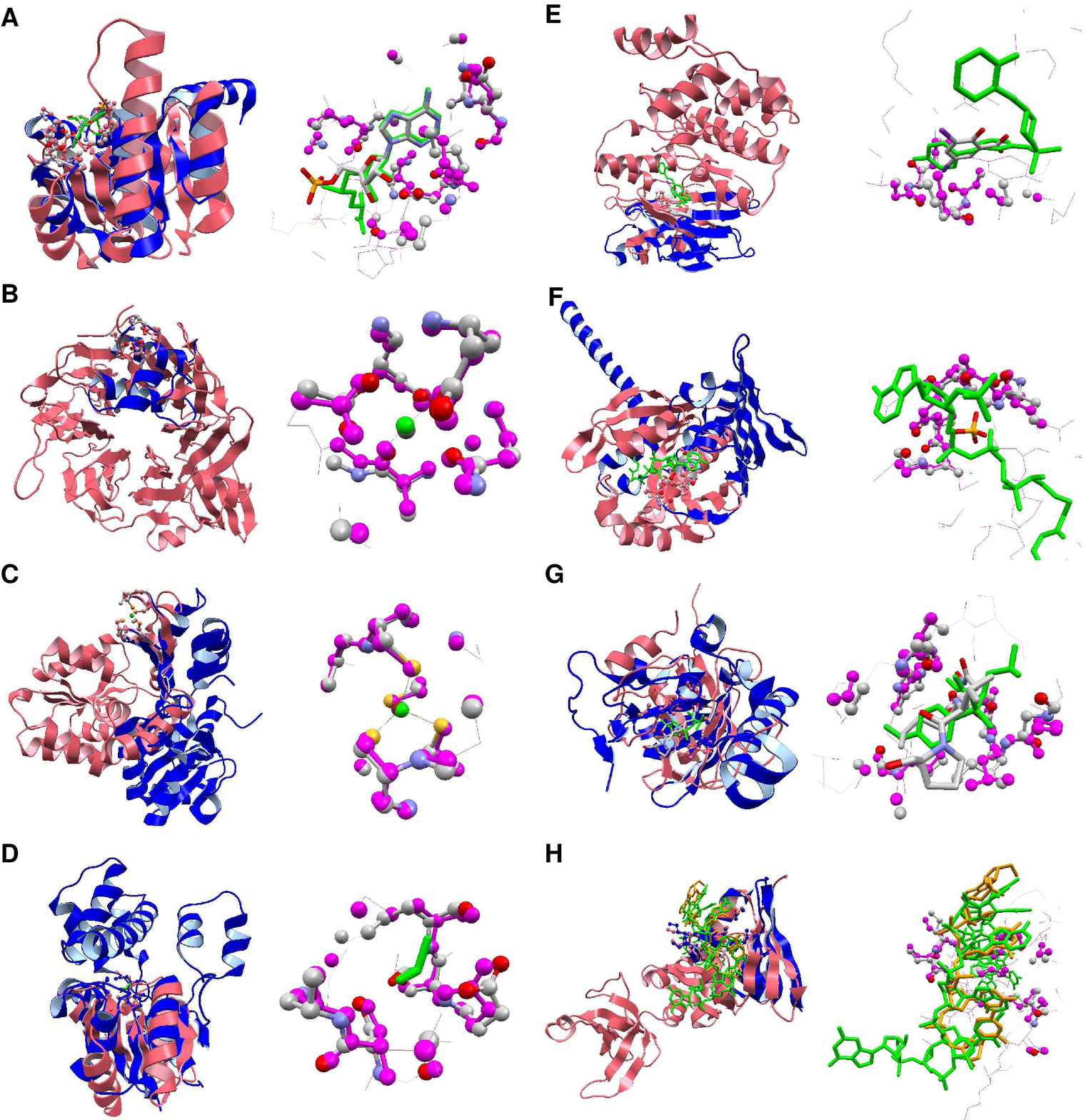}
  \caption{\figdiffoldlegend}
\end{figure*}
\section*{Discussion}
\label{discussion}
From the result of the exhaustive all-against-all comparison, 
we were able to obtain an extensive list of ligand binding site similarities
irrespective of sequence homology or global protein fold.
The similarity network uncovered very many cross-fold similarities as well 
as well-known ones. 
Although it is still not clear how many of these similarities are 
functionally relevant, it was often observed that different folds were 
superimposable to a significant extent when the alignment was based on the 
ligand binding sites (e.g., Figs. \figdifc{}, \figdiffold{}).
Aligning protein structures based on ligand binding sites 
(or functional sites in general) irrespective of sequence similarity, sequence 
order, and protein fold (as currently defined) may be a useful approach to 
elucidating the evolutionary history of fold changes \citep{Grishin2001b,KrishnaANDGrishin2004,AndreevaANDMurzin2006,Taylor2007,Goldstein2008,XieANDBourne2008}.

As was seen in the similarity network (Fig. \fignetwork), some links are 
based on the similarity of highly regular (secondary) structures which are 
found in many protein structures (e.g., Fig. \fignetwork{}B,F). 
While such similarities may not be  directly
related to any biochemical functions, they suggest that many ligand binding 
sites are based on combinations of some regular local structures.
It is known that a relatively small library of backbone fragments can accurately model tertiary structures of proteins \citep{KolodnyETAL2002}.
Consequently, the variety of contiguous fragments recurring in ligand binding 
sites is also limited as far as backbone structure is concerned. 
\citet{FriedbergANDGodzik2005} found similarities across different protein 
folds including those involved in various zinc-finger motifs and Rossmann-like 
folds, as shown in this study.
They also showed significant correlations between similarity of fragments 
and that of protein functions. This observation is consistent with the present 
results in that it suggests that specific combinations of fragments encode 
specific functions.
To apply the GIRAF method to functional annotations, however,
it is preferable to discriminate functionally relevant similarities 
from purely structural similarities. 

Some of the short-comings of simple pairwise comparison may be overcome by 
the complete-linkage clustering analysis of similar binding sites, which 
allowed us to define precise structural motifs. 
It should be stressed that defining reliable motifs requires redundancy 
in the PDB \citep{WangikarETAL2003}, otherwise it would be more difficult to 
distinguish recurring structures from incidental matches. 
These motifs may be useful for defining structural templates for 
efficient motif matching \citep{WallaceETAL1997}. 
Despite the diversity of binding sites and 
their similarities, most motifs were found to be confined within single 
families or superfamilies, and they were also found to be highly specific to 
particular ligands. Thus, these motifs may be helpful for annotating 
putative functions of proteins, especially, of structural genomics targets.
 
In conclusion, the development of an extremely efficient search method (GIRAF)
 to detect local structural similarities made it possible to conduct the first 
exhaustive all-against-all comparison of all ligand binding sites in all 
the known protein structures. We identified a number of well-defined 
structural motifs, enumerated many non-trivial similarities.
While exhaustive pairwise comparisons are useful for detecting
weak and possibly partial similarities between ligand binding sites, 
the significance of such matches may not be immediately obvious because 
some of them may be based on ubiquitous regular structures. 

Meanwhile, complete-linkage clusters of ligand binding sites are useful 
for identifying
functionally relevant binding site structures, but they may neglect partial 
but significant matches. Therefore, these two approaches, exhaustive pairwise 
comparison and motif matching, are complementary to each other, 
and hence the combination thereof may be helpful for more reliable annotations 
of proteins with unknown functions. These approaches may be further 
supplemented by other existing fold and/or 
sequence-based methods \citep{StandleyETAL2008b,XieANDBourne2008}.
The present method can be also applied to a whole protein structure 
(not limited to its predefined ligand binding sites) to find potential ligand 
binding sites \citep{KinjoANDNakamura2007}.
In this way, we are currently annotating all structural genomics targets \citep{TargetDB}.
We also plan to make this method available as a web service 
so that structural biologists can routinely search for ligand binding sites 
of their interest.

\section*{Experimental Procedures}
\label{sec:method}

\subsection*{The GIRAF method}
The details of the original GIRAF method has been published elsewhere \citep{KinjoANDNakamura2007}.
In this study, an improved version of GIRAF was used 
for conducting the all-against-all comparison. 
The improvement includes more sensitive geometric indexing with atomic 
composition around each reference set, simplified SQL expressions, and 
parallelization (A.R.K. and H.N., unpublished). 

\subsection*{All-against-all comparison}
Ligand binding sites were extracted from PDBML files as described in Results.
A ligand is defined as those entities that are annotated neither 
as ``polypeptide(L)'' with more than 24 amino acid residues 
nor ``water'' in the entity category. That is, a ligand can be polypeptide 
shorter than 25 residues, DNA, RNA, polysaccharides (sugars), lipids, 
metal ions, iron-sulfur clusters, or any other small molecules. 
However, ligands with more than 1000 atoms were discarded.
The all-against-all comparison was carried out on a cluster machine 
consisting of 20 nodes of 8-core processors (Intel Xeon 3.2 GHz). 
The whole computation was finished within approximately 60 hours.

\subsection*{Clusters of similar ligand binding sites}

To obtain complete-linkage clusters, we first constructed a single-linkage 
network based on a pre-defined P-value threshold. Then this network was 
decomposed into connected components. Each component was then broken into 
finer components by imposing a more stringent P-value threshold. 
This decomposition was iterated until P-value threshold reached $10^{-100}$. 
Then bottom-up complete-linkage was iteratively applied to each connected 
component, the result of which was then combined into an upper component 
(previously determined with a higher P-value threshold). 
This bottom-up process was terminated when P-value threshold, $10^{-15}$, 
was reached.
Each cluster was defined as a structural motif for the ligand binding sites.

\subsection*{Analysis of networks and structural motifs}

To annotate thus obtained structural motifs with the SCOP \citep{SCOP} codes, 
we used the parsable file of SCOP  (version 1.73).
When an analysis involved SCOP codes,
those PDB entries whose SCOP classification has not yet been determined were 
ignored. Each SCOP SCCS code was assigned to a ligand binding site as described
by others \citep{GoldANDJackson2006}. When a site resides at an interface of 
multiple domains, multiple SCCS codes were assigned to the site. 
Two or more binding sites are said to share the same fold 
(or family, superfamily, etc.) if the intersection of their SCCS code sets 
is not empty.
The SCCS code assigned to a structural motif was defined as the union of 
all the SCCS codes found in the corresponding cluster members.
We used only the seven main SCOP classes 
(all-$\alpha$ [a], all-$\beta$ [b], $\alpha/\beta$ [c], $\alpha + \beta$ [d], 
multi-domain [e], membrane and cell surface proteins and peptides [f],
and small proteins [g]).
The figures of alignments (Figs. \figdifc{} and \figdiffold{}) were created with jV version 3 \citep{KinoshitaANDNakamura2004} using the PDBML-extatom files produced by GIRAF.
The network figures (Fig. \fignetwork{}B-F) were created with the Tulip 
software (http://www.tulip-software.org/). 

\bibliographystyle{cell}
\bibliography{refs,mypaper}

\begin{thebibliography}{}

\bibitem[Altschul {\it et~al.}, 1997]{AltschulETAL1997}
Altschul, S.~F., Madden, T.~L., Schaffer, A.~A., Zhang, J., Zhang, Z., Miller,
  W., \& Lipman, D.~L. (1997).
\newblock Gapped blast and {PSI}-blast: A new generation of protein database
  search programs.
\newblock Nucleic Acids Res. {\em 25}, 3389--3402.

\bibitem[Andreeva \& Murzin, 2006]{AndreevaANDMurzin2006}
Andreeva, A. \& Murzin, A.~G. (2006).
\newblock Evolution of protein fold in the presence of functional constraints.
\newblock Curr. Opin. Struct. Biol. {\em 16}, 399--408.

\bibitem[Bachhawat {\it et~al.}, 2005]{1ZES}
Bachhawat, P., Swapna, G.~V., Montelione, G.~T., \& Stock, A.~M. (2005).
\newblock Mechanism of activation for transcription factor {PhoB} suggested by
  different modes of dimerization in the inactive and active states.
\newblock Structure {\em 13}, 1353--1363.

\bibitem[Barber {\it et~al.}, 1992]{2TMD}
Barber, M.~J., Neame, P.~J., Lim, L.~W., White, S., \& Matthews, F.~S. (1992).
\newblock Correlation of {X}-ray deduced and experimental amino acid sequences
  of trimethylamine dehydrogenase.
\newblock J. Biol. Chem. {\em 267}, 6611--6619.

\bibitem[Berman {\it et~al.}, 2007]{wwPDB}
Berman, H., Henrick, K., Nakamura, H., \& Markley, J.~L. (2007).
\newblock The worldwide {Protein Data Bank} ({wwPDB}): ensuring a single,
  uniform archive of {PDB} data.
\newblock Nucleic Acids Res. {\em 35}, D301--D303.

\bibitem[Berry \& {Phillips Jr.}, 1998]{1ZIN}
Berry, M.~B. \& {Phillips Jr.}, G.~N. (1998).
\newblock Crystal structures of \emph{{Bacillus} stearothermophilus} adenylate
  kinase with bound {Ap5A}, {Mg$^{2+}$} {Ap5A}, and {Mn$^{2+}$} {Ap5A} reveal
  an intermediate lid position and six coordinate octahedral geometry for bound
  {Mg$^{2+}$} and {Mn$^{2+}$}.
\newblock Proteins {\em 32}, 276--288.

\bibitem[Beuth {\it et~al.}, 2005]{2ATW}
Beuth, B., Pennell, S., Arnvig, K.~B., Martin, S.~R., \& Taylor, I.~A. (2005).
\newblock Structure of a \emph{{Mycobacterium} tuberculosis} {NusA-RNA}
  complex.
\newblock EMBO J. {\em 24}, 3576--3587.

\bibitem[Brakoulias \& Jackson, 2004]{BrakouliasANDJackson2004}
Brakoulias, A. \& Jackson, R.~M. (2004).
\newblock Towards a structural classification of phosphate binding sites in
  protein-nucleotide complexes: an automated all-against-all structural
  comparison using geometric matching.
\newblock Proteins {\em 56}, 250--260.

\bibitem[Carvalho {\it et~al.}, 2007]{2CCL}
Carvalho, A.~L., Dias, F. M.~V., Nagy, T., Prates, J. A.~M., Proctor, M.~R.,
  Smith, N., Bayer, E.~A., Davies, G.~J., Ferreira, L. M.~A., Romao, M.~J.,
  Fontes, C. M. G.~A., \& Gilbert, H.~J. (2007).
\newblock Evidence for a dual binding mode of dockerin modules to cohesins.
\newblock Proc. Natl. Acad. Sci. USA {\em 104}, 3089--3094.

\bibitem[Chen {\it et~al.}, 2004]{TargetDB}
Chen, L., Oughtred, R., Berman, H.~M., \& Westbrook, J. (2004).
\newblock {TargetDB}: a target registration database for structural genomics
  projects.
\newblock Bioinformatics {\em 20}, 2860--2862.

\bibitem[Dias {\it et~al.}, 2007]{2DFT}
Dias, M.~V., Faim, L.~M., Vasconcelos, I.~B., de~Oliveira, J.~S., Basso, L.~A.,
  Santos, D.~S., \& de~Azevedo, W.~F. (2007).
\newblock Effects of the magnesium and chloride ions and shikimate on the
  structure of shikimate kinase from \emph{{Mycobacterium} tuberculosis}.
\newblock Acta Crystallogr. F {\em 63}, 1--6.

\bibitem[Du {\it et~al.}, 2005]{2AXY}
Du, Z., Lee, J.~K., Tjhen, R., Li, S., Pan, H., Stroud, R.~M., \& James, T.~L.
  (2005).
\newblock Crystal structure of the first kh domain of human poly({C})-binding
  protein-2 in complex with a {C}-rich strand of human telomeric {DNA} at 1.7
  {{\AA}}.
\newblock J. Biol. Chem. {\em 280}, 38823--38830.

\bibitem[Friedberg \& Godzik, 2005]{FriedbergANDGodzik2005}
Friedberg, I. \& Godzik, A. (2005).
\newblock Connecting the protein structure universe by using sparse recurring
  fragments.
\newblock Structure {\em 13}, 1213--1224.

\bibitem[Garcia-Molina {\it et~al.}, 2002]{DB_complete}
Garcia-Molina, H., Ullman, J.~D., \& Widom, J. (2002).
\newblock {\em Database Systems: The Complete Book} (Prentice Hall: Upper
  Saddle River, NJ, U. S. A.).

\bibitem[Gold \& Jackson, 2006]{GoldANDJackson2006}
Gold, N.~D. \& Jackson, R.~M. (2006).
\newblock Fold independent structural comparisons of protein-ligand binding
  sites for exploring functional relationships.
\newblock J. Mol. Biol. {\em 355}, 1112--1124.

\bibitem[Goldstein, 2008]{Goldstein2008}
Goldstein, R.~A. (2008).
\newblock The structure of protein evolution and the evolution of protein
  structure.
\newblock Curr. Opin. Struct. Biol. {\em 18}, 170--177.

\bibitem[Gonzalez {\it et~al.}, 2004]{1W2D}
Gonzalez, B., Schell, M.~J., Letcher, A.~J., Veprintsev, D.~B., Irvine, R.~F.,
  \& Williams, R.~L. (2004).
\newblock Structure of a human inositol 1,4,5-trisphosphate 3-kinase: substrate
  binding reveals why it is not a phosphoinositide 3-kinase.
\newblock Mol. Cell {\em 15}, 689--701.

\bibitem[Grishin, 2001a]{Grishin2001b}
Grishin, N.~V. (2001a).
\newblock Fold change in evolution of protein structures.
\newblock J. Struct. Biol. {\em 134}, 167--185.

\bibitem[Grishin, 2001b]{Grishin2001}
Grishin, N.~V. (2001b).
\newblock {KH} domain: one motif, two folds.
\newblock Nucleic Acids Res. {\em 29}, 638--643.

\bibitem[Guilloteau {\it et~al.}, 2002]{1LQY}
Guilloteau, J.~P., Mathieu, M., Giglione, C., Blanc, V., Dupuy, A., Chevrier,
  M., Gil, P., Famechon, A., Meinnel, T., \& Mikol, V. (2002).
\newblock The crystal structures of four peptide deformylases bound to the
  antibiotic actinonin reveal two distinct types: a platform for the
  structure-based design of antibacterial agents.
\newblock J. Mol. Biol. {\em 320}, 951--962.

\bibitem[Gutteridge \& Thornton, 2005]{GutteridgeANDThornton2005}
Gutteridge, A. \& Thornton, J.~M. (2005).
\newblock Understanding nature's catalytic toolkit.
\newblock Trends Biochem. Sci. {\em 30}, 622--629.

\bibitem[Herter {\it et~al.}, 2005]{1Z8G}
Herter, S., Piper, D.~E., Aaron, W., Gabriele, T., Cutler, G., Cao, P., Bhatt,
  A.~S., Choe, Y., Craik, C.~S., Walker, N., Meininger, D., Hoey, T., \&
  Austin, R.~J. (2005).
\newblock Hepatocyte growth factor is a preferred in vitro substrate for human
  hepsin, a membrane-anchored serine protease implicated in prostate and
  ovarian cancers.
\newblock Biochem. J. {\em 390}, 125--136.

\bibitem[Hoff {\it et~al.}, 2006]{2H4H}
Hoff, K.~G., Avalos, J.~L., Sens, K., \& Wolberger, C. (2006).
\newblock Insights into the sirtuin mechanism from ternary complexes containing
  {NAD}$^{+}$ and acetylated peptide.
\newblock Structure {\em 14}, 1231--1240.

\bibitem[Ikura {\it et~al.}, 2008]{IkuraETAL2008}
Ikura, T., Kinoshita, K., \& Ito, N. (2008).
\newblock A cavity with an appropriate size is the basis of the ppiase
  activity.
\newblock Protein Eng. Des. Sel. {\em 21}, 83--89.

\bibitem[{Joint Center for Structural Genomics}, 2006]{2G1U}
{Joint Center for Structural Genomics} (2006).
\newblock Crystal structure of (tm1088a) from \emph{{Thermotoga} maritima} at
  1.50 {{\AA}} resolution.
\newblock PDB entry 2G1U.

\bibitem[Jones \& Thornton, 2004]{JonesANDThornton2004}
Jones, S. \& Thornton, J.~M. (2004).
\newblock Searching for functional sites in protein structures.
\newblock Curr. Opin. Struct. Biol. {\em 8}, 3--7.

\bibitem[Kawabata, 2003]{Kawabata2003}
Kawabata, T. (2003).
\newblock Matras: a program for protein {3D} structure comparison.
\newblock Nucleic Acids Res. {\em 31}, 3367--3369.

\bibitem[Kawabata \& Nishikawa, 2000]{KawabataANDNishikawa2000}
Kawabata, T. \& Nishikawa, K. (2000).
\newblock Protein tertiary structure comparison using the markov transition
  model of evolution.
\newblock Proteins {\em 41}, 108--122.

\bibitem[Kinjo \& Nakamura, 2007]{KinjoANDNakamura2007}
Kinjo, A.~R. \& Nakamura, H. (2007).
\newblock Similarity search for local protein structures at atomic resolution
  by exploiting a database management system.
\newblock BIOPHYSICS {\em 3}, 75--84.
\newblock doi:10.2142/biophysics.3.75.

\bibitem[Kinoshita \& Nakamura, 2004]{KinoshitaANDNakamura2004}
Kinoshita, K. \& Nakamura, H. (2004).
\newblock {eF}-site and {PDBjViewer}: database and viewer for protein
  functional sites.
\newblock Bioinformatics {\em 20}, 1329--1330.

\bibitem[Kinoshita {\it et~al.}, 1999]{KinoshitaETAL1999}
Kinoshita, K., Sadanami, K., Kidera, A., \& Go, N. (1999).
\newblock Structural motif of phosphate-binding site common to various protein
  superfamilies: all-against-all structural comparison of
  protein-mononucleotide complexes.
\newblock Protein Eng. {\em 12}, 11--14.

\bibitem[Kobayashi \& Go, 1997]{KobayashiANDGo1997}
Kobayashi, N. \& Go, N. (1997).
\newblock {ATP} binding proteins with different folds share a common
  {ATP}-binding structural motif.
\newblock Nat. Struct. Biol. {\em 4}, 6--7.

\bibitem[Kolodny {\it et~al.}, 2002]{KolodnyETAL2002}
Kolodny, R., Koehl, P., Guibas, L., \& Levitt, M. (2002).
\newblock Small libraries of protein fragments model native protein structures
  accurately.
\newblock J. Mol. Biol. {\em 323}, 297--307.

\bibitem[Krishna \& Grishin, 2004]{KrishnaANDGrishin2004}
Krishna, S.~S. \& Grishin, N.~V. (2004).
\newblock Structurally analogous proteins do exist!.
\newblock Structure {\em 12}, 1125--1127.

\bibitem[Krishna {\it et~al.}, 2003]{KrishnaETAL2003}
Krishna, S.~S., Majumdar, I., \& Grishin, N.~V. (2003).
\newblock Structural classification of zinc fingers: survey and summary.
\newblock Nucleic Acids Res. {\em 31}, 532--550.

\bibitem[Krissinel \& Henrick, 2004]{SSM}
Krissinel, E. \& Henrick, K. (2004).
\newblock Secondary-structure matching ({SSM}), a new tool for fast protein
  structure alignment in three dimensions.
\newblock Acta Crystallogr. D {\em 60}, 2256--2268.

\bibitem[Laronde-Leblanc {\it et~al.}, 2005]{1ZAR}
Laronde-Leblanc, N., Guszczynski, T., Copeland, T., \& Wlodawer, A. (2005).
\newblock Autophosphorylation of \emph{{Archaeoglobus} fulgidus} {Rio2} and
  crystal structures of its nucleotide-metal ion complexes.
\newblock FEBS J. {\em 272}, 2800--2810.

\bibitem[Lee {\it et~al.}, 2007]{LeeETAL2007}
Lee, D., Redfern, O., \& Orengo, C. (2007).
\newblock Predicting protein function from sequence and structure.
\newblock Nat. Rev. Mol. Cell Biol. {\em 8}, 995--1005.

\bibitem[Minai {\it et~al.}, 2008]{MinaiETAL2008}
Minai, R., Matsuo, Y., Onuki, H., \& Hirota, H. (2008).
\newblock Method for comparing the structures of protein ligand-binding sites
  and application for predicting protein-drug interactions.
\newblock Proteins {\em 72}, 367--381.

\bibitem[Murzin {\it et~al.}, 1995]{SCOP}
Murzin, A.~G., Brenner, S.~E., Hubbard, T., \& Chothia, C. (1995).
\newblock {SCOP}: A structural classification of proteins database for the
  investigation of sequences and structures.
\newblock J. Mol. Biol. {\em 247}, 536--540.

\bibitem[Nagano {\it et~al.}, 2002]{NaganoETAL2002}
Nagano, N., Orengo, C.~A., \& Thornton, J.~M. (2002).
\newblock One fold with many functions: the evolutionary relationships between
  {TIM} barrel families based on their sequences, structures and functions.
\newblock J. Mol. Biol. {\em 321}, 741--765.

\bibitem[Orengo {\it et~al.}, 1994]{OrengoETAL1994}
Orengo, C.~A., Jones, D.~T., \& Thornton, J.~M. (1994).
\newblock Protein superfamilies and domain superfolds.
\newblock Nature {\em 372}, 631--634.

\bibitem[Pattabhi {\it et~al.}, 2004]{1UHB}
Pattabhi, V., Syed~Ibrahim, B., \& Shamaladevi, N. (2004).
\newblock Trypsin activity reduced by an autocatalytically produced
  nonapeptide.
\newblock J. Biomol. Struct. Dyn. {\em 21}, 737--744.

\bibitem[Polacco \& Babbitt, 2006]{PolaccoANDBabbitt2006}
Polacco, B.~J. \& Babbitt, P.~C. (2006).
\newblock Automated discovery of {3D} motifs for protein function annotation.
\newblock Bioinformatics {\em 22}, 723--730.

\bibitem[Porter {\it et~al.}, 2004]{PorterETAL2004}
Porter, C.~T., Bartlett, G.~J., \& Thornton, J.~M. (2004).
\newblock {The Catalytic Site Atlas}: a resource of catalytic sites and
  residues identified in enzymes using structural data.
\newblock Nucleic Acids Res. {\em 32}, D129--D133.

\bibitem[Ridder {\it et~al.}, 1997]{1AQ6}
Ridder, I.~S., Rozeboom, H.~J., Kalk, K.~H., Janssen, D.~B., \& Dijkstra, B.~W.
  (1997).
\newblock Three-dimensional structure of {L}-2-haloacid dehalogenase from
  \emph{{Xanthobacter} autotrophicus} {GJ10} complexed with the
  substrate-analogue formate.
\newblock J. Biol. Chem. {\em 272}, 33015--33022.

\bibitem[Rognan, 2007]{Rognan2007}
Rognan, D. (2007).
\newblock Chemogenomic approaches to rational drug design.
\newblock Br. J. Pharmacol. {\em 152}, 38--52.

\bibitem[Shulman-Peleg {\it et~al.}, 2004]{Shulman-PelegETAL2004}
Shulman-Peleg, A., Nussinov, R., \& Wolfson, H.~J. (2004).
\newblock Recognition of functional sites in protein structures.
\newblock J. Mol. Biol. {\em 339}, 607--633.

\bibitem[Standley {\it et~al.}, 2008]{StandleyETAL2008b}
Standley, D.~M., Toh, H., \& Nakamura, H. (2008).
\newblock Functional annotation by sequence-weighted structure alignments:
  Statistical analysis and case studies from the {Protein} 3000 structural
  genomics project in {Japan}.
\newblock Proteins {\em 72}, 1333--1351.

\bibitem[Stark \& Russell, 2003]{StarkETAL2003}
Stark, A.~Sunyaev, S. \& Russell, R.~B. (2003).
\newblock A model for statistical significance of local similarities in
  structure.
\newblock J. Mol. Biol. {\em 326}, 1307--1316.

\bibitem[Stoll {\it et~al.}, 1997]{5GRT}
Stoll, V.~S., Simpson, S.~J., Krauth-Siegel, R.~L., Walsh, C.~T., \& Pai, E.~F.
  (1997).
\newblock Glutathione reductase turned into trypanothione reductase: structural
  analysis of an engineered change in substrate specificity.
\newblock Biochemistry {\em 36}, 6437--6447.

\bibitem[Tari {\it et~al.}, 1997]{1AQ2}
Tari, L.~W., Matte, A., Goldie, H., \& Delbaere, L.~T. (1997).
\newblock {Mg$^{2+}$-Mn$^{2+}$} clusters in enzyme-catalyzed
  phosphoryl-transfer reactions.
\newblock Nat. Struct. Biol. {\em 4}, 990--994.

\bibitem[Tari {\it et~al.}, 1996]{TariETAL1996}
Tari, L.~W., Mattte, A., Pugazhenthi, U., Goldie, H., \& Delbaere, L.~T.
  (1996).
\newblock Snashot of an enzyme reaction intermediate in the structure of the
  {ATP}-{Mg$^{2+}$}-oxalate ternary complex of \emph{Escherichia coli} {PEP}
  carboxykinase.
\newblock Nat. Struct. Biol. {\em 3}, 355--363.

\bibitem[Taylor, 2007]{Taylor2007}
Taylor, W.~R. (2007).
\newblock Evolutionary transitions in protein fold space.
\newblock Curr. Opin. Struct. Biol. {\em 17}, 354--361.

\bibitem[Wallace {\it et~al.}, 1997]{WallaceETAL1997}
Wallace, A.~C., Borkakoti, N., \& Thornton, J.~M. (1997).
\newblock {TESS}: A geometric hashing algorithm for deriving {3D} coordinate
  templates for searching structural databases. application to enzyme active
  sites.
\newblock Protein Sci. {\em 6}, 2308--2323.

\bibitem[Wangikar {\it et~al.}, 2003]{WangikarETAL2003}
Wangikar, P.~P., Tendulkar, A., Ramya, S., Mali, D.~N., \& Sarawagi, S. (2003).
\newblock Functional sites in protein families uncovered via an objective and
  automated graph theoretic approach.
\newblock J. Mol. Biol. {\em 326}, 955--978.

\bibitem[Watts \& Strogatz, 1998]{WattsANDStrogatz1998}
Watts, D.~J. \& Strogatz, S. (1998).
\newblock Collective dynamics of `small-world' networks.
\newblock Nature {\em 393}, 440--442.

\bibitem[Westbrook {\it et~al.}, 2005]{PDBML}
Westbrook, J., Ito, N., Nakamura, H., Henrick, K., \& Berman, H.~M. (2005).
\newblock {PDBML}: the representation of archival macromolecular structure data
  in {XML}.
\newblock Bioinformatics {\em 21}, 988--992.

\bibitem[Whitlow {\it et~al.}, 1999]{1QB1}
Whitlow, M., Arnaiz, D.~O., Buckman, B.~O., Davey, D.~D., Griedel, B.,
  Guilford, W.~J., Koovakkat, S.~K., Liang, A., Mohan, R., Phillips, G.~B.,
  Seto, M., Shaw, K.~J., Xu, W., Zhao, Z., Light, D.~R., \& Morrissey, M.~M.
  (1999).
\newblock Crystallographic analysis of potent and selective factor {Xa}
  inhibitors complexed to bovine trypsin.
\newblock Acta Crystallogr., D {\em 55}, 1395--1404.

\bibitem[Wolfson \& Rigoutsos, 1997]{GeometricHashing}
Wolfson, H.~J. \& Rigoutsos, I. (1997).
\newblock Geometric hashing: An overview.
\newblock IEEE Comput. Sci. Eng. {\em 4}, 10--21.

\bibitem[Wu {\it et~al.}, 2006]{2NYU}
Wu, H., Dong, A., Zeng, H., Loppnau, P., Weigelt, J., Sundstrom, M.,
  Arrowsmith, C.~H., Edwards, A.~M., Bochkarev, A., \& Plotnikov, A.~N. (2006).
\newblock The crystal structure of human {FtsJ} homolog 2 (\emph{{E}. coli})
  protein in complex with {AdoMet}.
\newblock PDB entry 2NYU.

\bibitem[Xiao {\it et~al.}, 2004]{1TXV}
Xiao, T., Takagi, J., Coller, B.~S., Wang, J.~H., \& Springer, T.~A. (2004).
\newblock Structural basis for allostery in integrins and binding to
  fibrinogen-mimetic therapeutics.
\newblock Nature {\em 432}, 59--67.

\bibitem[Xie \& Bourne, 2008]{XieANDBourne2008}
Xie, L. \& Bourne, P.~E. (2008).
\newblock Detecting evolutionary relationships across existing fold space,
  using sequence order-independent profile-profile alignments.
\newblock Proc. Natl. Acad. Sci. {USA} {\em 105}, 5441--5446.

\end{thebibliography}

\section*{Acknowledgments}

The authors thank Kengo Kinoshita, Motonori Ota and Hiroyuki Toh for helpful 
discussion, and Daron M. Standley for critically reading the manuscript.
This work was supported by a grant-in-aid from Institute for Bioinformatics 
Research and Development, Japan Science and Technology Agency (JST). 
H. N. was supported by Grant-in-Aid for Scientific Research (B) No. 20370061 
from Japan Society for the Promotion of Science (JSPS).

%%%%%
\end{document}